\documentclass[sigconf]{acmart}
\usepackage{xcolor}

\definecolor{RevisionColor}{HTML}{002FA7}

\usepackage{xcolor}
\usepackage{makecell}
\usepackage{array}

\AtBeginDocument{%
  }

\copyrightyear{2026}
\acmYear{2026}
\setcopyright{cc}
\setcctype{by}
\acmConference[CHI '26]{Proceedings of the 2026 CHI Conference on Human Factors in Computing Systems}{April 13--17, 2026}{Barcelona, Spain}
\acmBooktitle{Proceedings of the 2026 CHI Conference on Human Factors in Computing Systems (CHI '26), April 13--17, 2026, Barcelona, Spain}
\acmDOI{10.1145/3772318.3791330}
\acmISBN{979-8-4007-2278-3/2026/04}

\begin{teaserfigure}
\includegraphics[width=\linewidth, keepaspectratio]{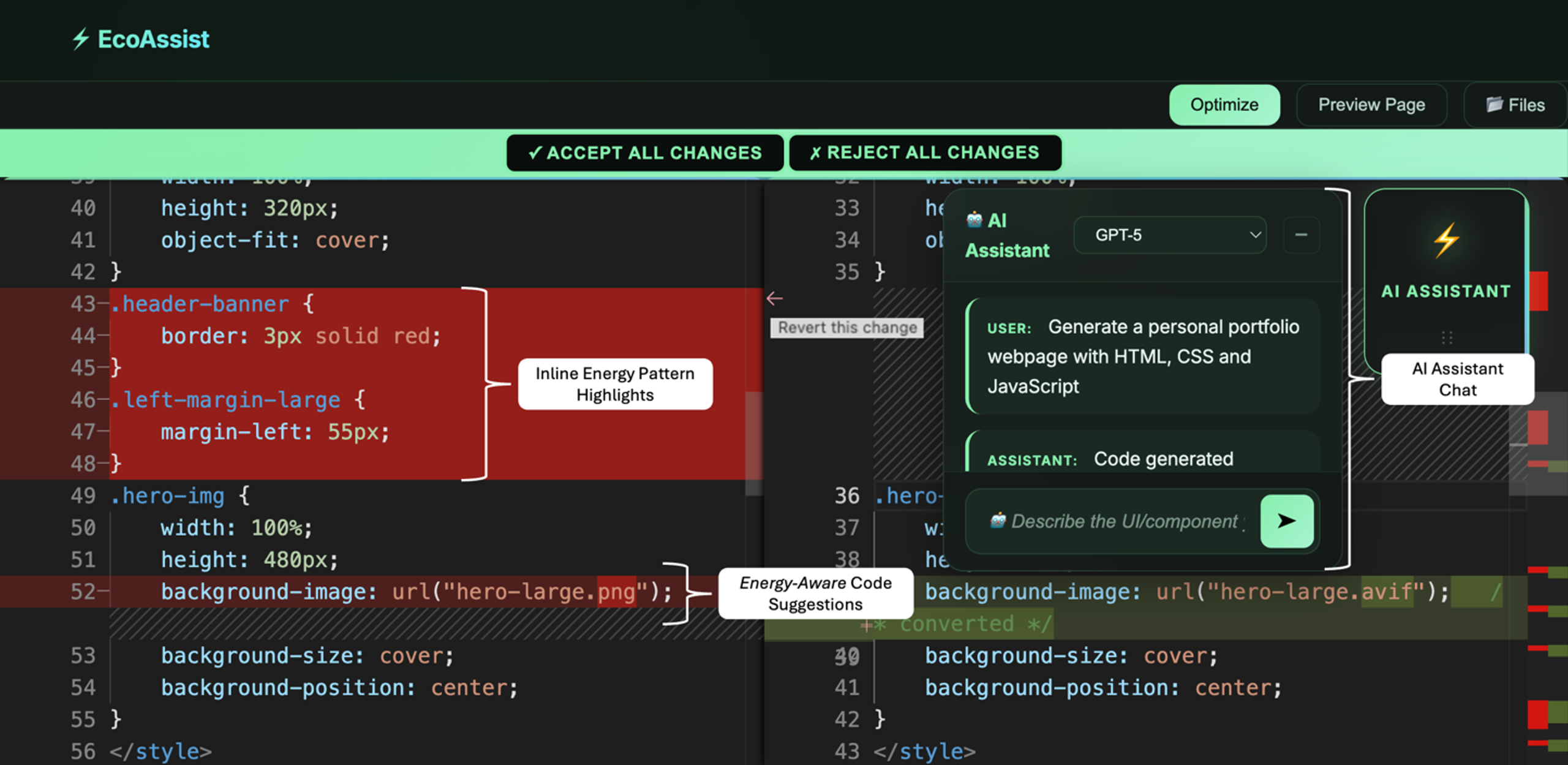}
\caption{EcoAssist is an AI-assisted tool for energy-aware front-end development. It supports developers by (1) highlighting energy-intensive patterns, (2) suggesting optimized alternatives, and (3) enabling selective acceptance of changes. Integrated directly into the coding workflow, EcoAssist promotes energy awareness while helping produce lower-energy front-end code.}
\Description{A screenshot of the EcoAssist editor interface showing sections of code flagged as energy-intensive. Optimized alternatives are suggested inline and can be accepted or rejected by the developer.}
\end{teaserfigure}

\begin{document}

\title[EcoAssist: Embedding Sustainability into AI-Assisted Frontend Development]{EcoAssist: Embedding Sustainability into AI-Assisted Frontend Development}


\author{André Barrocas}
\email{andre.barrocas@tecnico.ulisboa.pt}
\orcid{0000-0001-9804-3129}
\affiliation{%
  \department{ITI / LARSyS, Instituto Superior Técnico}
  \institution{University of Lisbon}
  \city{Lisbon}
  \country{Portugal}
}

\author{Nuno Jardim Nunes}
\email{nunojnunes@tecnico.ulisboa.pt}
\orcid{0000-0002-2498-0643}
\affiliation{%
  \department{ITI / LARSyS, Instituto Superior Técnico}
  \institution{University of Lisbon}
  \city{Lisbon}
  \country{Portugal}
}

\author{Valentina Nisi}
\orcid{0000-0002-8051-3230}
\email{valentina.nisi@tecnico.ulisboa.pt}
\affiliation{%
  \department{ITI / LARSyS, Instituto Superior Técnico}
  \institution{University of Lisbon}
  \city{Lisbon}
  \country{Portugal}
}

\author{Nikolas Martelaro}
\email{nikmart@cmu.edu}
\orcid{0000-0002-1824-0243}
\affiliation{%
    \department{Human-Computer Interaction Institute}
    \institution{Carnegie Mellon University}
    \city{Pittsburgh}
    \state{Pennsylvania}
    \country{USA}
}

\renewcommand{\shortauthors}{Barrocas et al.}
\begin{abstract}
Frontend code, replicated across millions of page views, consumes significant energy and contributes directly to digital emissions. Yet current AI coding assistants, such as GitHub Copilot and Amazon CodeWhisperer, emphasize developer speed and convenience, with energy impact not yet a primary focus. At the same time, existing energy-focused guidelines and metrics have seen limited adoption among practitioners, leaving a gap between research and everyday coding practice. To address this gap, we introduce EcoAssist, an energy-aware assistant integrated into an IDE that analyzes AI-generated frontend code, estimates its energy footprint, and proposes targeted optimizations. We evaluated EcoAssist through benchmarks of 500 websites and a controlled study with 20 developers. Results show that EcoAssist reduced per-website energy by ~13–16\% on average, increased developers’ awareness of energy use, and maintained developer productivity. This work demonstrates how energy considerations can be embedded directly into AI-assisted coding workflows, supporting developers as they engage with energy implications through actionable feedback.
\end{abstract}




\maketitle

\section{Introduction}
Modern web frontend development frequently overlooks energy efficiency, resulting in resource-intensive websites with high energy consumption \cite{kalliola2025challenges,vemuri2021greening}. A typical web page visit emits approximately 0.4 grams of CO$_2$e, and cumulatively, the Information and Communication Technology (ICT) sector is estimated to contribute ~1.8–2.8\% of global greenhouse gas emissions, with adjusted analysis placing the range between ~2.1–3.9\%, a footprint comparable to that of the aviation sector \cite{wholegrain2025,freitag2021real}. Current web performance initiatives, such as Google’s Core Web Vitals, emphasize  loading speed and responsiveness for a better user experience \cite{gaddam2025optimizing,wehner2022vital} but do not address the carbon emissions associated with inefficient code. Recent efforts in the software community have produced formal sustainability guidelines for web design, notably the W3C’s Web Sustainability Guidelines (WSG) released in 2023 \cite{w3c2023wsg,wimalasena2025toward}, and established best practices, such as the Sustainable Web Manifesto’s principles for creating environmentally responsible and energy-efficient digital products \cite{lecorney2024review}. 

Frameworks such as the Sustainable Web Design Model (SWDM) and tools like CO2.js \cite{greenweb2025co2js,kalliola2025challenges} now enable estimation of digital carbon emissions based on data transfer and energy sources. Yet these resources remain largely outside everyday development workflows. Studies show that many web developers have low awareness of sustainable coding practices and rarely use sustainability tools during development \cite{hulleberg2023awareness}. In fast-paced production environments, environmental concerns are often overshadowed by feature delivery, leaving a gap between high-level sustainability goals and the actual code being written \cite{noman2022exploratory}.

Simultaneously, AI-powered coding assistants are being adopted in frontend development \cite{liang2024large}. Systems like GitHub Copilot \cite{github2023copilot} are now widely used to generate frontend code snippets (e.g. HTML, CSS, and JavaScript) on the fly, significantly accelerating development by enabling a ``vibe coding'' style of quickly prototyping with AI-suggested code \cite{cui2024productivity,imai2022github,peng2023impact}. However, current assistants optimize primarily for functionality and convenience, rather than energy efficiency or carbon emissions \cite{cursaru2024controlled,rani2025can}. They may suggest solutions that involve large libraries, unoptimized images, or wasteful computations, inadvertently increasing a webpage’s overall energy consumption \cite{alizadeh2025language,ashraf2025energy}. Critically, these tools provide no feedback on the energy-efficiency of generated code, and metrics such as emissions, runtime energy consumption, and performance overhead remain invisible to developers \cite{islam2025evaluating,solovyeva2025ai}. As a result, even well-meaning developers can remain unaware of energy-intensive patterns embedded in AI-produced solutions \cite{cappendijk2025exploration}. 

In this paper, we address the need to integrate energy awareness into AI-assisted frontend development. We present EcoAssist, an energy-aware frontend coding assistant integrated directly into an IDE that provides energy-oriented sustainability metrics and recommendations to reduce the energy consumption of AI-generated code. It suggests optimized alternatives, such as removing unused code, converting images to efficient formats, and minimizing JavaScript file sizes, with the aim of reducing the energy consumption of frontend code without compromising functionality. 

To inform the design of EcoAssist, we reviewed prior work on AI-assisted coding and sustainable web development to understand how developers work with AI code assistants and where energy-intensive patterns emerge. Our analysis of prior work highlights three key gaps:
\begin{enumerate}
    \item Developers and AI coding assistants rarely initiate discussions about energy efficiency or carbon emissions~\cite{cappendijk2025exploration}.
    \item Inefficient patterns in AI-generated code often remain unnoticed without explicit cues~\cite{islam2025evaluating}.
    \item Developers lack mechanisms to validate energy impact during development, unlike the immediate functional feedback provided by browsers~\cite{hulleberg2023awareness}.
\end{enumerate}
EcoAssist embeds energy analysis directly into existing developer workflows, evaluating and annotating frontend code as it is written. We evaluated EcoAssist through benchmarks of 500 websites and a controlled developer study. The results show that EcoAssist reduces energy consumption, improves performance, and raises developer awareness of energy-related implications, all without slowing down the development process. Our findings demonstrate that integrating energy feedback into AI-assisted coding workflows can effectively connect sustainability guidelines with everyday coding decisions.
The contributions of this paper are: (1) an empirical analysis of AI-assisted coding focused on energy consumption, identifying how and why current AI workflows (e.g., with GitHub Copilot) overlook sustainability considerations; (2) the design and implementation of EcoAssist, an energy-aware coding assistant that exposes energy-related issues directly within the development environment; and (3) an evaluation demonstrating that EcoAssist produces more energy-efficient frontend code than AI assistants alone, while raising developer awareness of energy considerations.

\section{Related Work} \label{relatedwork}

Our research builds upon four strands of prior work: (i) Assessing Web Sustainability, (ii) Energy-Aware User Behavior, (iii) Energy Efficiency in Development Tools, and (iv) Energy Efficiency of LLM-Generated Code.

\subsection{Evaluating Web Sustainability}
Over the past decade, sustainable computing has garnered attention through the development of standards, measurement frameworks, and advocacy efforts \cite{danushi2025carbon,isoiec2024sci}. The W3C’s Web Sustainability Guidelines (WSG) \cite{w3c2023wsg} established a conceptual foundation for environmentally responsible web design. Similarly, the Green Software Foundation has advanced the Software Carbon Intensity (SCI) specification, which recently achieved ISO standardization \cite{greensoftware2024sci,isoiec2024sci}. The SCI defines a method for calculating the rate of carbon emissions per unit of work performed by software (e.g., grams of CO$_2$ equivalent per user request), offering a standardized way to compare and monitor software sustainability \cite{greensoftware2024sci}. 

Frameworks such as the Sustainable Web Design Model (SWDM) \cite{sustainablewebdesign2024} provide a structured method for estimating digital emissions by translating key inputs, including page weight and data transfer, into carbon emissions. The SWDM is an open-source model that estimates emissions by considering energy usage across data centers, networks, and end-user devices \cite{sustainablewebdesign2024,zhu2025enhancing}. Complementary initiatives, such as recent empirical studies, have emphasized the growing share of software systems in global energy use and highlighted the importance of energy-aware web design \cite{khrouf2025energy}. Furthermore, empirical investigations have revealed recurring inefficiencies in web development, such as frequent client-side polling, oversized images, and unused scripts or styles, all of which increase energy consumption and network costs \cite{rani2024energy}. More efficient practices, such as event-driven updates and serving images at appropriate resolutions, can lower the energy footprint of user-facing applications \cite{rani2024energy}. Additionally, prior research has demonstrated the potential to reduce frontend energy by replacing standard JavaScript interactive components with static alternatives, resulting in approximately 5\% energy savings across real-world webpages without compromising interactivity \cite{fouquet2022jsrehab}. 

Building on this line of work, subsequent studies of JavaScript-based applications have revealed significant differences in energy consumption between implementations, with consumption varying substantially across different libraries and network conditions \cite{miettinen2010analysis,selakovic2016performance}. Empirical analysis of JavaScript performance issues reveals that inefficient API usage is the most prevalent cause of energy-intensive operations, with most energy optimizations requiring only minimal code changes \cite{tong2023comparative}. 

EcoAssist builds on this work by providing integrated energy optimization capabilities for frontend development workflows. Our system continuously evaluates frontend code during the development process, automatically identifying and flagging energy-intensive patterns such as excessive DOM manipulations, inefficient CSS selectors, resource-heavy animations, and suboptimal JavaScript operations that contribute to increased CPU usage and energy drain. Beyond detection, EcoAssist actively suggests alternative, concrete, energy-efficient implementations as developers write code or review AI-generated snippets. This integrated approach enables developers to make informed decisions, reducing the environmental footprint of web applications through optimized resource use, decreased processing overhead, and improved energy efficiency across user devices.

\subsection{Energy-Aware User Behavior}

Beyond measurement approaches, energy-aware computing practices have evolved to incorporate behavioral interventions and design methodologies that promote environmental awareness and energy-conscious usage \cite{blevis2007sustainable,disalvo2010mapping,silberman2014next}. Digital minimalism frameworks encourage users to adopt lighter browsing habits, such as reducing video streaming quality during peak energy hours and choosing text-based content over media-rich alternatives \cite{newport2019digital}. Similarly, sustainable user experience (UX) design principles advocate for simplified interfaces that require fewer computational resources \cite{issa2022sustainable,nayak2021green}. Implementing techniques such as lazy loading, progressive enhancement, and content prioritization helps minimize energy consumption during user interactions \cite{frick2016designing}.

Educational initiatives have also emerged to bridge the knowledge gap between digital consumption and environmental impact \cite{hajj2024environmental,jaradat2024integrating}. Organizations like the Green Web Foundation have developed educational resources and verification systems that help developers and organizations understand the ecological implications of their hosting choices, providing transparency about which providers use renewable energy \cite{greenweb2025co2js,greenweb2025}. These approaches demonstrate how integrating energy awareness into daily digital habits can create a widespread, positive impact \cite{hajj2024environmental}.

However, even when people try to use digital services in an energy-conscious way, their influence is limited by how efficient those services are in the first place. If a system is built inefficiently, careful user behavior cannot meaningfully offset that. This shifts the core responsibility to developers, whose design and implementation choices set the system’s underlying energy impact. EcoAssist addresses this gap by highlighting energy concerns as developers work, reinforcing energy-aware behavior during creation. 

\subsection{Energy Efficiency in Development Tools}

Reducing the environmental impact of digital systems depends on the choices made by developers \cite{danushi2025carbon,young2025empowering}. Several tools already support energy-efficient coding indirectly through performance optimization \cite{connolly2025software,cruz2017performance}. Chrome DevTools \cite{chromedevtools2025} and Firefox Profiler \cite{mozilla2025firefoxprofiler}, for instance, offer detailed runtime analysis of long tasks, memory usage, and repaint events, with some support for performance signals correlated with energy on select devices \cite{pourghassemi2019if}. Projects like GreenHub \cite{matalonga2019greenhub}, a collaborative platform that crowdsources real-world energy data from Android devices, and PowerAPI \cite{fieni2024powerapi}, a modular software-defined power meter for attributing consumption to processes, enable fine-grained empirical monitoring of energy usage across devices.

Researchers have developed tools such as Lacuna and Muzeel \cite{kupoluyi2021muzeel,malavolta2023javascript} that automatically detect and eliminate JavaScript dead code from web applications, demonstrating that the removal of unused code significantly reduces loading times and network transfer costs. These studies indicate that dead code elimination approaches can reduce the computational overhead of frontend applications while maintaining functional equivalence \cite{kupoluyi2021muzeel,malavolta2023javascript}. More recently, LLM-based frameworks such as SysLLMatic \cite{peng2025sysllmatic} have extended these ideas beyond rule-based detection. SysLLMatic \cite{peng2025sysllmatic} combines LLM reasoning with system profiling to optimize performance, energy, and throughput across complex software stacks. While effective, this approach is not specifically tailored to the needs of frontend development, which limits its integration into modern workflows \cite{bakal2025experience,peng2023impact,perumal2025role}. Despite recent advances, many energy inefficiencies in frontend development (e.g., unoptimized images, inefficient libraries, inefficient rendering patterns) still remain unnoticed in practice due to their subtle nature \cite{dornauer2023web}. Existing sustainability and energy metrics can quantify these costs but rarely integrate seamlessly into development workflows \cite{danushi2025carbon}. Additionally, automated techniques such as image compression and code minification still run outside the development workflow, offering optimisation only after code is written rather than during implementation \cite{chen2025need,ilager2025green}. As a result, developers must leave their coding environments to consult external tools, and no feedback is provided during the software development process \cite{danushi2025carbon,oyedeji2024integrating}.

EcoAssist addresses these limitations by embedding energy insights into the development tools. It automatically inspects AI-suggested frontend snippets, flags energy-intensive patterns, and proposes optimized alternatives grounded in established frameworks. Thus, EcoAssist focuses on two complementary areas: (1) extending sustainable web development and green code research into the emerging practice of AI-assisted “vibe coding,” where current assistants prioritize rapid development while overlooking energy efficiency and its environmental consequences, and (2) integrating energy considerations into developer workflows to help balance productivity with environmental impact. 

\subsection{Energy Efficiency of LLM-Generated Code}

Recent progress in large pretrained models, such as Large Language Models (LLMs) \cite{brown2020language,vaswani2017attention} and Large Multimodal Models (LMMs) \cite{achiam2023gpt,radford2021learning}, has created opportunities to improve efficiency in software development \cite{chen2021evaluating,li2022competition,wang2021codet5}. For instance, these systems can streamline repetitive coding tasks and assist in selecting lighter-weight frameworks, indirectly reducing developer effort and system overhead \cite{fried2022incoder}. However, their contributions have largely focused on productivity and convenience, with energy-efficiency impacts of the code itself remaining less visible \cite{verdecchia2023systematic}. For instance, AI coding assistants, such as GitHub Copilot \cite{github2023copilot}, enhance productivity and speed, but typically overlook energy costs \cite{bird2022taking}. Even when prompted, LLMs like Code Llama rarely produce energy-efficient implementations \cite{cui2024productivity,cursaru2024controlled}. Prior work suggests that the performance characteristics of AI-generated code are mixed and context-dependent, with some studies reporting inefficiencies or limitations in handling implicit constraints, while others find that AI-generated code can score higher on maintainability and documentation and lower on cyclomatic complexity than human-written code \cite{eltabakh2024quality}. Other work shows that LLM-produced programs can include unnecessary complexity and verbosity, increasing computational overhead in some settings \cite{azeem2025ai}.

More recently, LLM-based frameworks such as EffiLearner \cite{huang2024effilearner} have advanced automated optimization of LLM-generated code. EffiLearner \cite{huang2024effilearner} utilizes runtime feedback to refine LLM-generated code iteratively, achieving reductions of up to 87\% in execution time and 90\% in memory usage. However, this approach is designed for general software optimization and is not tailored to the specific requirements of frontend development (e.g., HTML, CSS, JavaScript). More recently, Green-Code \cite{ilager2025green} has shown that LLM inference energy can be reduced by 23–50\% through dynamic early exits, with a VS Code prototype suggesting the feasibility of IDE-level feedback. Yet, its focus is on making the model itself more efficient, not on the sustainability of the frontend code generated by AI assistants \cite{ilager2025green}. 

EcoAssist fills these gaps by adding frontend-specific energy feedback to the AI-assisted coding process. Unlike prior approaches, which either improve LLM inference efficiency or optimize code without focusing on frontend concerns, EcoAssist operates within AI-assisted development, helping developers produce more efficient frontend code.

\section{EcoAssist}
\label{ecoassist}

EcoAssist is an energy-aware coding assistant that embeds energy awareness reasoning directly into AI-assisted frontend development. Rather than treating energy efficiency as an after-the-fact audit, EcoAssist integrates it into the vibe coding workflow. As developers generate or write code, the system analyzes energy costs, proposes optimized alternatives, and highlights the energy impact. EcoAssist learns from an offline pipeline that pairs real energy measurements with concrete code changes, and it then distills this knowledge into a lightweight runtime assistant that operates inside the editor.

\subsection{Design Goals}

To formulate EcoAssist design goals, we revisit established guidelines on sustainable web development \cite{sustainablewebdesign2024,w3c2023wsg}, recent advances in energy-aware software systems \cite{chan2020investigating,de2023comparison,pockstaller2023comparing}, and prior work on designing human-AI interactions in coding contexts \cite{peng2023impact}. We highlight two overarching goals that frame EcoAssist’s contribution: (G1) extending sustainable web development and green code research into AI-assisted “vibe coding,” and (G2) integrating energy awareness and efficiency comparison directly into developer workflows.

\subsubsection{Extending Sustainable Web Development into AI-Assisted Coding}

Research in sustainable web development has introduced tools for profiling and optimization, such as Chrome DevTools, Firefox Profiler, and platforms like GreenHub and PowerAPI \cite{chromedevtools2025,fieni2024powerapi,matalonga2019greenhub,mozilla2025firefoxprofiler}. While useful, these solutions typically act after code is written. At the same time, the rise of AI coding assistants like GitHub Copilot has encouraged rapid “vibe coding,” where oversized assets, redundant dependencies, and inefficient DOM operations slip in without energy-efficiency feedback \cite{dakhel2023github,imai2022github}. Yet current tools remain disconnected from this fast-paced, AI-driven workflow \cite{ilager2025green,kupoluyi2021muzeel,malavolta2023javascript}. Therefore, our first design goal is to bridge this gap by embedding energy checks and optimizations into the fast-paced context of frontend coding. Unlike external assessment platforms, EcoAssist integrates directly into the AI-assisted development loop, analyzing energy implications of generated code and proposing optimized alternatives. By doing so, it ensures that sustainability considerations are addressed alongside productivity, transforming energy efficiency from a late-stage check into a primary concern in everyday coding practice \cite{dakhel2023github}. This proactive approach avoids the limitations of purely rule-based, after-the-fact optimizations, ensuring that inefficient patterns are prevented at the point of generation rather than corrected only once they have already been introduced.

\subsubsection{Integrating Energy Awareness into Developer Workflows}

Most energy-oriented tools operate outside developers’ everyday environments, reporting raw metrics like bundle size or estimated emissions without clear guidance on what to do next \cite{danushi2025carbon,pathania2025calculating}. While useful for measurement, they rarely integrate into IDEs or directly support decision-making, leaving developers to balance energy-efficiency against usability and delivery speed on their own \cite{venters2023sustainable}. Even newer frameworks improve efficiency but mostly flag issues, offering little help for developers to understand trade-offs like energy cost versus visual quality or delays from deferring scripts \cite{huang2024effilearner,ilager2025green,peng2025sysllmatic,peng2023impact}. Building on these ideas, our second design goal is to integrate energy awareness into AI-assisted processes. This involves presenting optimizations in terms of energy impact (e.g., estimated joules saved) while coding or generating frontend code with AI. By providing concrete alternatives, such as replacing unoptimized image formats, delaying the loading of non-critical scripts, or reducing redundant dependencies, EcoAssist turns sustainability tools from passive monitors into active collaborators that help developers balance productivity, maintainability, and environmental impact.

\subsection{System Overview}

At a high level, EcoAssist comprises two connected phases: an offline training pipeline and an online runtime optimizer. The offline pipeline is responsible for building the dataset that grounds the system in empirical evidence by linking code changes to measured energy outcomes. We assembled a corpus of 500 websites from the Kaggle Phishing Website dataset for fine-tuning \cite{huntingdata11_2023_phishing}.

For each website in this corpus, EcoAssist begins with baseline energy profiling on both server and client, using Powermetrics \cite{apple2023powermetrics} to capture system-level energy consumption and Playwright \cite{microsoft2025playwright} to drive automated browser interactions. The webpage is then optimized with established libraries, re-profiled under the same conditions, and compared against the original to quantify differences in energy use and performance. The resulting before-and-after pairs are stored as structured training examples.

\begin{figure*}[t]
\fboxsep=0mm
\includegraphics[width=\textwidth]{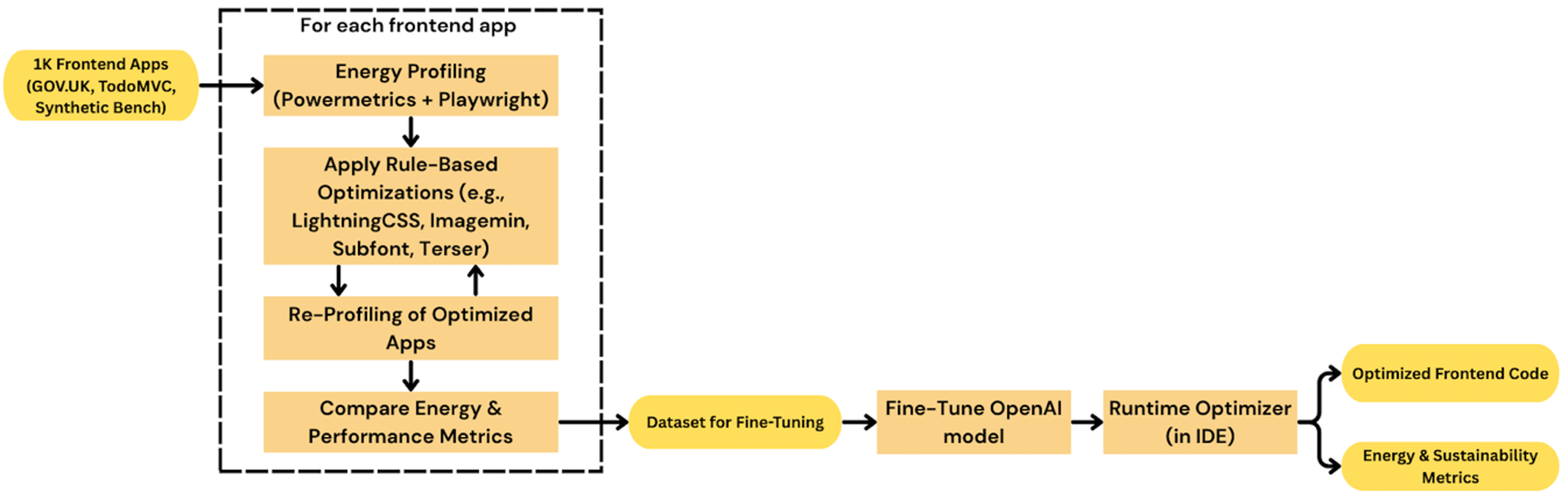}
\caption{Workflow shows offline training pipeline (left) generating fine-tuning dataset, which trains model (center) deployed in runtime optimizer (right)}
\Description{A flow diagram showing EcoAssist’s two-phase architecture: (1) the offline pipeline that generates training data to fine-tune the model, and (2) the online runtime optimizer that injects sustainability-aware suggestions directly within the developer’s IDE.}
\label{fig:fig-2}
\end{figure*}

From this training dataset, we fine-tune a general-purpose code model to recognize inefficient frontend patterns, propose edits that reduce energy consumption while maintaining functionality. The online phase exposes this capability directly within the developer’s IDE. As code is written or inserted by an AI assistant, EcoAssist inspects the snippet, produces an optimized alternative, and presents estimated energy savings. Figure \ref{fig:fig-2} illustrates the offline pipeline that generates training data and the runtime optimizer that delivers sustainability-aware guidance during frontend development.

\subsection{Offline Training Pipeline}

The goal of the offline pipeline is to ground EcoAssist’s runtime optimizer in real energy evidence. This phase is performed once and produces the fine-tuning dataset used to adapt a general-purpose language model into a energy-aware coding assistant.

We assembled a corpus of 500 websites drawn from the Phishing Website HTML Classification dataset on Kaggle for fine-tuning the model \cite{huntingdata11_2023_phishing}. These webpages were used only for training and were not reused in the benchmark described in the Results section. Each website underwent a four-step pipeline. First, we conducted energy profiling to capture baseline consumption. On the server side, we used Powermetrics \cite{apple2023powermetrics} on Apple MacBook Air (M4, 16 GB RAM, macOS Sequoia 15.6.1), using Chrome to measure CPU and GPU energy during rendering and API responses. On the client side, we used Playwright \cite{microsoft2025playwright} to drive scripted browser interactions and measured the associated CPU and GPU energy with Powermetrics. 

Second, we applied rule-based optimizations using established tools: LightningCSS \cite{parcel2023lightningcss} for stylesheet minification, Terser \cite{terser2018} for JavaScript minification, HTML-Minifier-Terser \cite{htmlminifier2020} for HTML cleanup, Subfont \cite{subfont2017} for font subsetting, Critical \cite{osmani2015critical} for inlining above-the-fold styles, and Imagemin \cite{imagemin2013} and Squoosh \cite{squoosh2018} for image compression and AVIF transcoding. We also evaluated a set of widely used build and optimization frameworks, including esbuild \cite{wallace2025esbuild}, SWC \cite{swc2025}, Vite \cite{vite2025}, Parcel \cite{parcel2025}, and SVGO \cite{svgo2025}. These tools can each reduce energy consumption in specific contexts. For example, build-time bundlers tend to be most effective for JavaScript-heavy pages, while SVGO is particularly useful for graphics optimization. We tested them individually, measured their impact, and incorporated only those optimizations that consistently produced measurable energy improvements. Third, we paired the original and optimized versions of the code, recording the exact transformation applied and the measured change in energy. Finally, we assembled these before/after pairs into a supervised dataset, ensuring that the model was exposed only to optimization patterns that reliably reduced energy. Using this dataset, we fine-tuned OpenAI’s GPT-4o-mini to specialize in frontend energy optimization, training it to recognize inefficient code patterns and propose replacements grounded in empirically validated improvements. While independent optimization tools can save energy, an energy-aware assistant can apply them faster and with less effort.

\subsection{Online Runtime Optimizer}

Once the model has been fine-tuned, EcoAssist provides a lightweight runtime optimizer that runs inside the developer’s coding environment. It evaluates the energy use of the original code snippet, applies the model’s improvements, and then re-evaluates the optimized version using the same measurement procedure as in training. The process runs quickly and unobtrusively, allowing developers to receive feedback on energy reduction while coding.

This runtime optimizer is integrated into the IDE, where it inspects and improves frontend snippets. When invoked, it sends the snippet to the fine-tuned GPT-4o-mini model, which returns a version that preserves structure, layout, and behavior while applying optimizations such as reducing redundant JavaScript, compressing oversized images, deferring non-critical scripts, or optimizing vector. Alongside the optimized code, EcoAssist computes a unified diff that highlights the exact changes, enabling developers to review and decide whether to accept or reject each optimization.

\subsection{User Interaction}

EcoAssist provides an integrated development environment (IDE) that enables developers to generate, edit, optimize, and preview frontend code while tracking the energy impact of code changes. It supports iterative workflows common in vibe-coding: developers can either generate frontend code with the AI assistant or write it manually. Users can choose which model to invoke (e.g., base GPT-4, GPT-4o, GPT-5, or Claude) to generate code and then open the preview to render the page. The preview feature allows developers to validate functionality and compare the optimized layout with the original one.

\begin{figure}[t]
\fboxsep=0mm
\includegraphics[width=\linewidth]{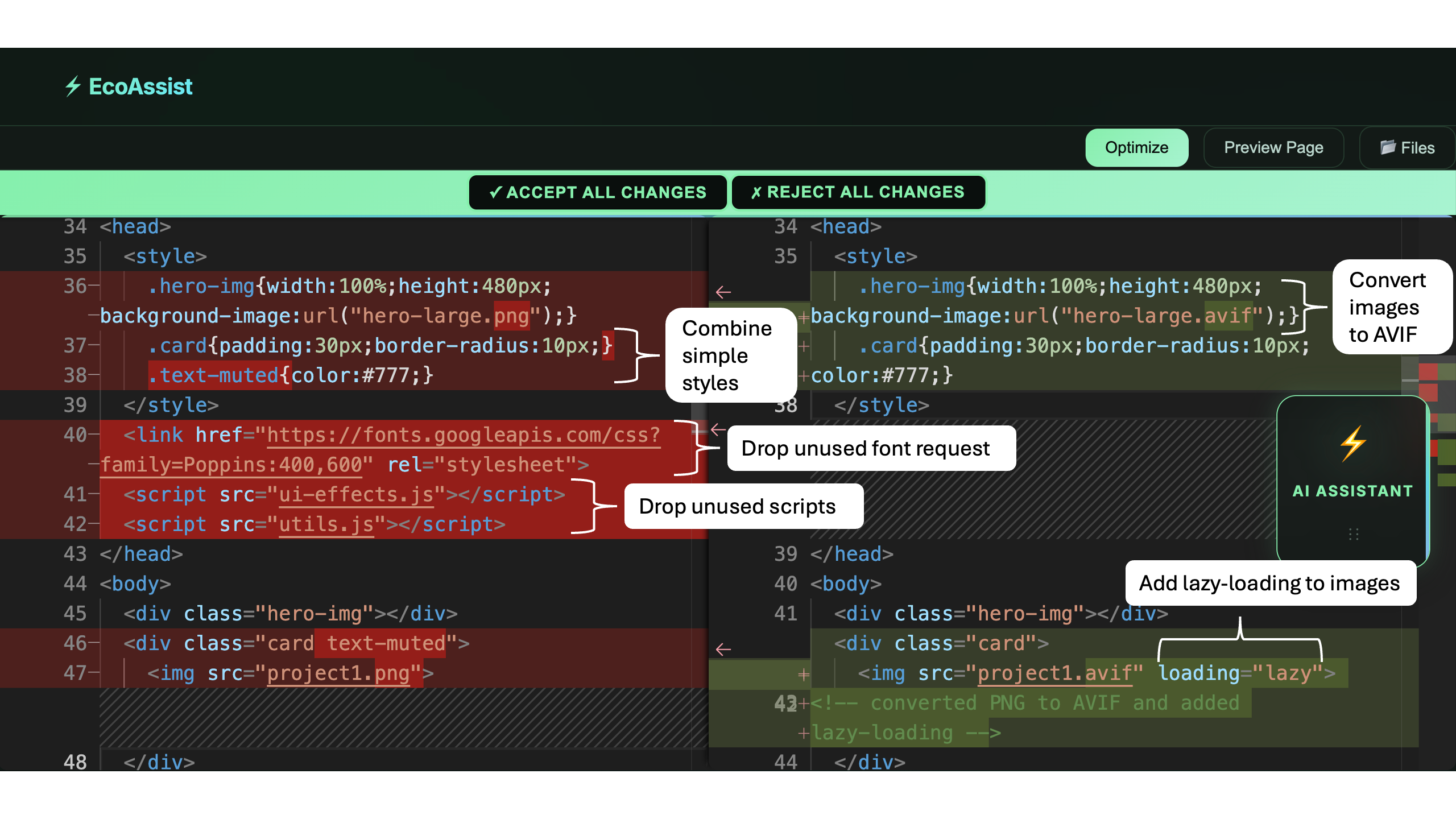}
\caption{EcoAssist’s inline optimization view, showing detected energy-intensive patterns (left) and optimized alternatives (right). Shown transformations include AVIF conversion, lazy-loading, simplified styles, and unused font/script removal, all applied through the accept/reject interface.}
\Description{A screenshot capturing EcoAssist in action, generating a sample personal portfolio website through an integrated AI assistant inside the IDE.}
\label{fig:fig-3}
\end{figure}

\begin{figure}[t]
\fboxsep=0mm
\includegraphics[width=\linewidth]{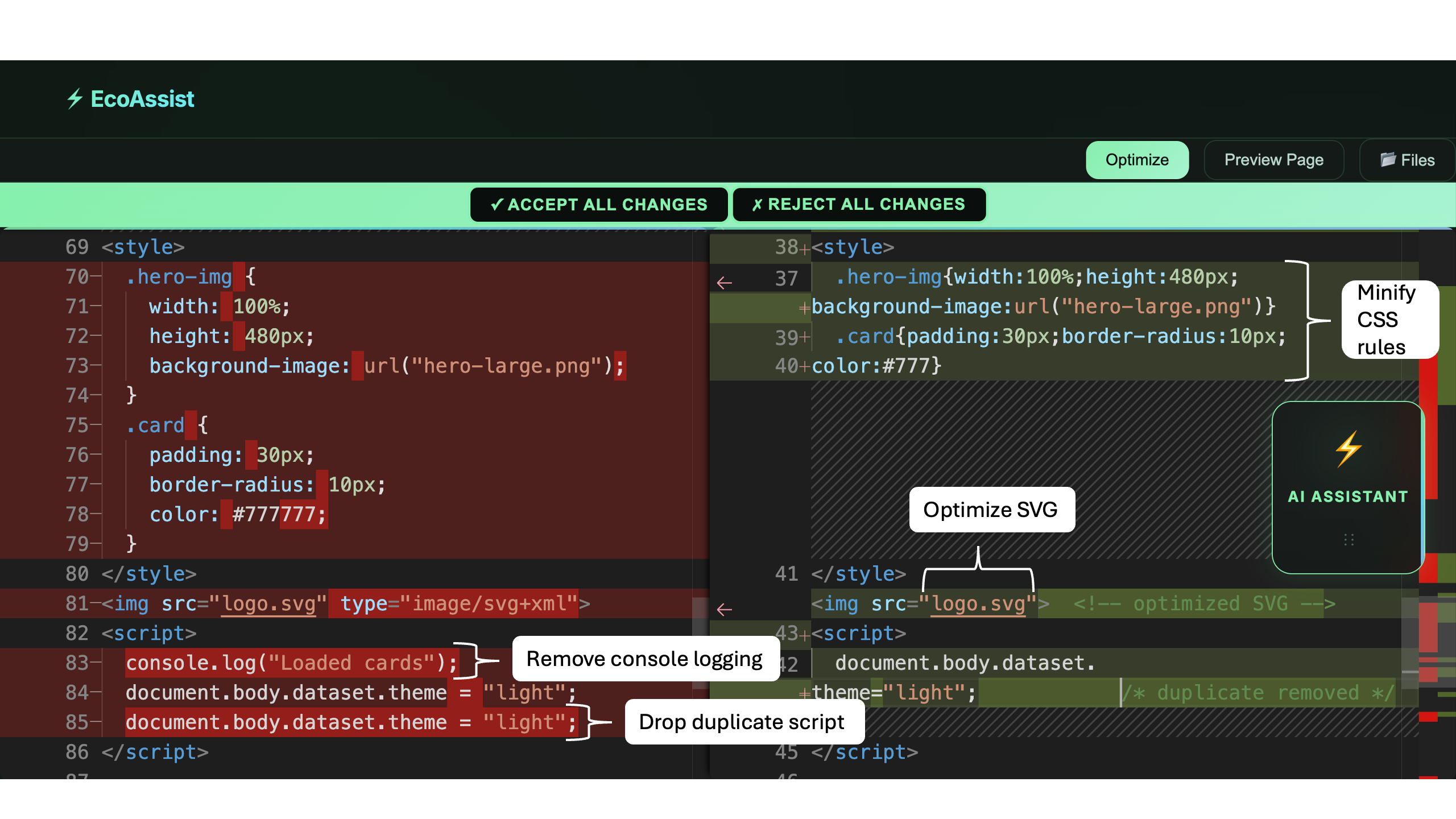}
\caption{EcoAssist’s diff view highlighting optimizations such as CSS minification, SVG optimization, and removal of redundant logging and scripts, all presented through the accept/reject interface.}
\Description{A side-by-side code editor view where the original code is displayed beside EcoAssist’s optimized version. Inline diffs highlight the changes, and buttons allow developers to reject transformations selectively.}
\label{fig:fig-4}
\end{figure}

\begin{figure}[t]
\fboxsep=0mm
\includegraphics[width=\linewidth]{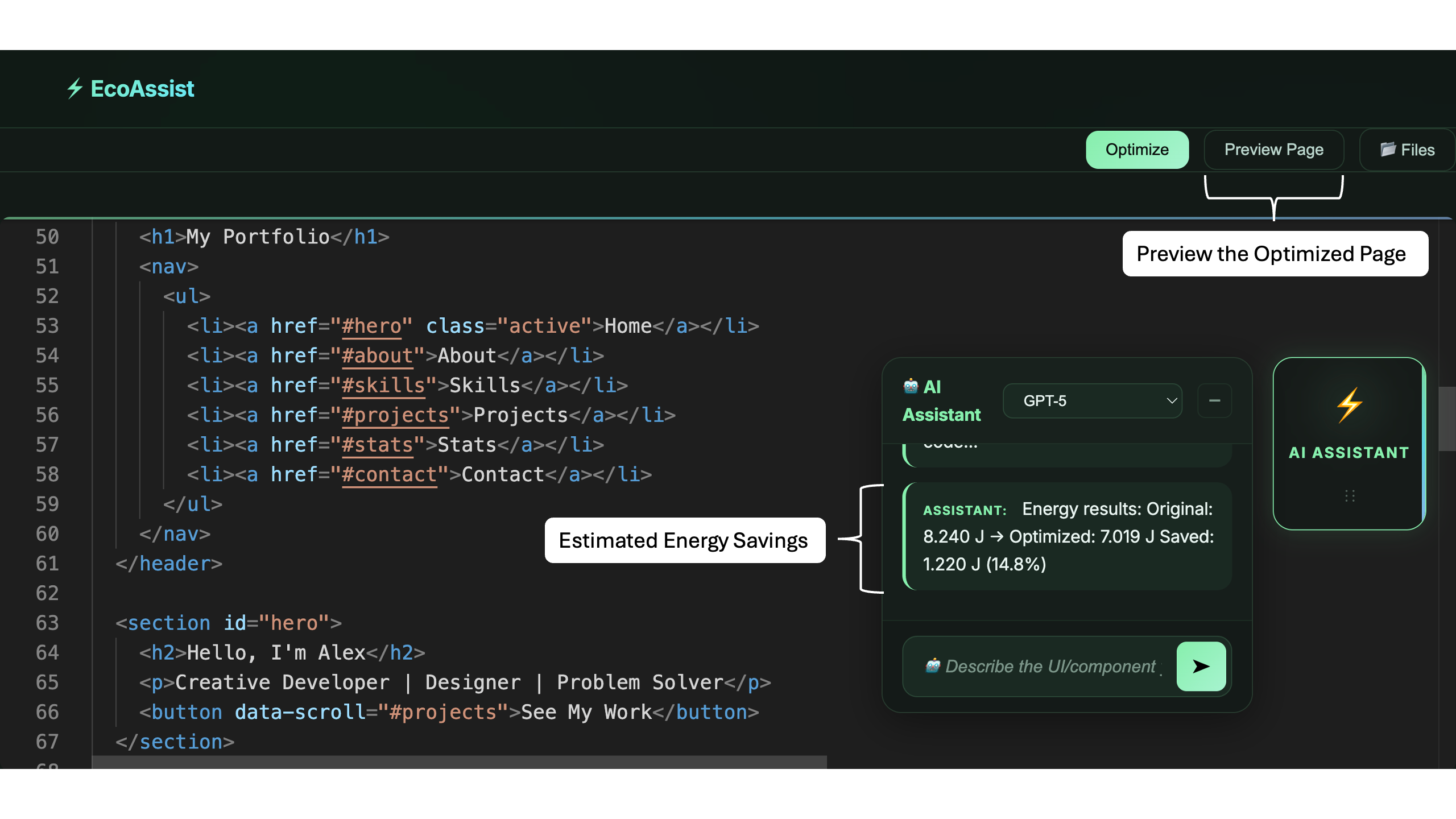}
\caption{EcoAssist compares energy use before and after optimization and lets developers preview the optimized page directly in the workflow.}
\Description{EcoAssist reporting energy savings by comparing energy usage before and after optimization. It shows both the absolute savings in joules and the relative reduction expressed as a percentage.}
\label{fig:fig-5}
\end{figure}

Optimization is handled directly within the environment. By invoking the optimizer, developers receive an improved version of their code. The editor then displays the original and optimized versions side by side, with inline diffs clearly highlighting transformations such as converting oversized images to AVIF, adding lazy-loading attributes to off-screen media, deferring or inlining non-critical scripts, removing unused CSS rules, and simplifying redundant DOM operations. These suggestions are presented as editable code changes grounded in EcoAssist’s training data, allowing developers to inspect, validate, and selectively apply them through the accept/reject interface illustrated in Figures \ref{fig:fig-3} and \ref{fig:fig-4}. In parallel, the AI assistant reports energy estimates for both versions, summarizing the expected savings in joules and as a percentage, as shown by Figure \ref{fig:fig-5}.

\section{Evaluation}\label{evaluation}

We evaluated EcoAssist to assess both its technical effectiveness in reducing energy consumption and its broader impact on everyday developer workflows. To capture these dimensions, our evaluation combined a comprehensive system benchmark of websites with a controlled user study that examined how developers interacted with the tool in practice.

\subsection{Methodology}

We designed our evaluation using a mixed-methods approach to capture both technical performance and developer experience. The first component of our assessment was a system benchmark designed to measure EcoAssist’s impact on energy consumption in a controlled setting. In this benchmark, we applied EcoAssist model to two disjoint datasets: 250 webpages generated by GPT-5 and 250 webpages sampled from the ``Phishing Website HTML Classification'' dataset \cite{huntingdata11_2023_phishing}, distinct from the 500 Kaggle pages used for fine-tuning. For each webpage, we measured energy consumption before and after optimization using our model, allowing us to quantify efficiency gains across diverse page types. Energy usage was recorded with Powermetrics \cite{apple2023powermetrics} on both the server and client sides, with Playwright \cite{microsoft2025playwright} automating Chrome to simulate user interactions on the client. 

To complement the benchmark study, we conducted a user study with twenty software developers. This user study was designed to assess whether EcoAssist’s optimizations preserved both functionality and visual aspects of the generated applications, while also examining its impact on usability, workload, and developers’ awareness of energy efficiency. Each participant completed a forty-five to sixty-minute session with the EcoAssist prototype, following a structured workflow that involved generating a complete personal portfolio website via the integrated AI assistant in EcoAssist (selecting the GPT-5 model), applying optimizations with our finetuned GPT-4o-mini model, and verifying that the optimized code preserved functionality while reporting energy savings. 

The participant set represented early-career developers with varying levels of experience. On average, participants reported 4.7 years of general coding experience (range 2--10) and 2.5 years of frontend development experience (range 0--8). This group represented technically competent developers with moderate frontend experience but limited familiarity with energy-aware coding practices. Most participants (18/20, 90.0\%) reported prior use of AI coding assistants such as Copilot or Cursor, while only a minority (2/20, 10.0\%) had never used such tools. Familiarity with sustainable web development was modest ($M = 2.3$ on a 4-point scale, $SD = 0.8$), and only two participants had previously worked with tools that directly measure energy or emissions. 

Data collection combined quantitative and qualitative measures. Quantitative data captured both technical outcomes and self-reported perceptions, while qualitative data captured participants’ reflections during the study. Developers were recruited through university mailing lists, and all sessions were conducted in person in a lab setting following a structured, facilitator-guided workflow. Technical outcomes included absolute and relative energy savings per task. All self-reported measures were collected through a structured post-study questionnaire. This questionnaire included three standardized instruments: (1) usability was assessed with the System Usability Scale (SUS) \cite{bangor2008empirical}; (2) workload was measured using the NASA Task Load Index (NASA-TLX) \cite{hart1988development}, focusing on mental, temporal, and frustration demands, as these dimensions are relevant for assessing whether EcoAssist introduces cognitive overhead or disrupts developers’ workflow; and (3) sustainability sensitivity was measured with an adapted version of the Energy Awareness concepts from Jagroep et al. \cite{jagroep2017awakening}. We created three items to assess whether participants understood their code’s energy use, could identify optimizations, and felt supported in making sustainable coding decisions. Qualitative insights were obtained from participants’ verbal reflections captured during the recorded user sessions, which followed a think-aloud protocol during the task and concluded with a short post-study questionnaire and a semi-structured interview. These materials were reviewed to identify common patterns and recurring observations about usability, sustainability reasoning, and interaction with EcoAssist. The combination of the benchmark and the developer sessions offered a comprehensive view of EcoAssist’s effectiveness in embedding energy awareness into AI-assisted frontend development.

\subsection{Results}

\subsubsection{System Benchmark}

\begin{figure*}[t]
\fboxsep=0mm
\includegraphics[width=\linewidth]{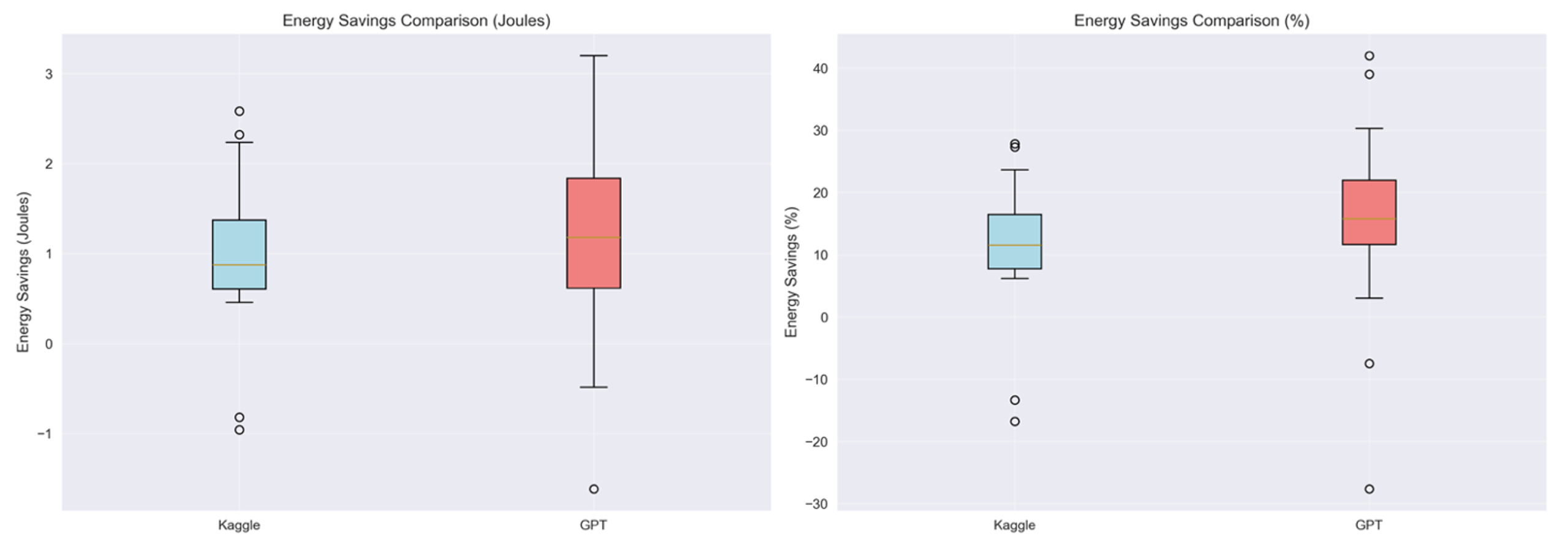}
\caption{Energy savings with EcoAssist model on Kaggle vs. GPT webpages, shown in joules (left) and percentage (right).}
\Description{A pair of box plots presenting the results of the benchmark. One showing energy savings in joules, the other displaying percentage improvements, comparing EcoAssist’s performance on AI-generated pages versus real-world Kaggle pages.}
\label{fig:fig-6}
\end{figure*}

We evaluated EcoAssist on a test set of 500 webpages (250 GPT-generated using the GPT-5 model and 250 real-world pages from Kaggle \cite{huntingdata11_2023_phishing}). Websites were selected using criteria of file size $>600$ KB and JavaScript intensity $\geq10$ DOM-affecting operations (measured via scripted instrumentation), ensuring the dataset represented real-world complex web applications with substantial energy optimization potential. Energy usage was measured in joules on a controlled testbed (Apple MacBook Air M4, 16 GB RAM, macOS Sequoia 15.6.1, using Chrome 139.0.7258.155 (arm64)) with Powermetrics \cite{apple2023powermetrics} and Playwright \cite{microsoft2025playwright}. Each page was profiled five times over a 4-second execution window with a 2-second cooldown period between runs to ensure thermal stability and measurement consistency. Results were reported as the mean across all runs. The cache was cleared, browser extensions were disabled, brightness was set to 50\%, and the device was connected to AC power. On average, EcoAssist reduced per-page energy consumption by 13.4\% (95\% CI [10.2, 16.6]). Improvements varied across datasets: GPT-generated pages showed larger reductions ($M = 15.4\%$, $SD = 13.4\%$) than Kaggle pages ($M = 11.5\%$, $SD = 9.8\%$), as shown by Figure \ref{fig:fig-6}. However, an independent two-sample Welch's t-test on per-page means found this difference not significant ($p = 0.238$), indicating that EcoAssist delivers broadly comparable optimization effectiveness across both AI-generated and real-world content types. Overall, optimizations were highly effective: 93\% of all pages exhibited reduced energy use, and 70\% achieved savings above 10\%. Performance outcomes complemented the energy savings. Median network transfer decreased by 8--10\%, and median code size (HTML/CSS/JS) was reduced by about 7\%, indicating that EcoAssist’s optimizations generally made pages both lighter and more efficient to process. These reductions support the conclusion that energy gains align with improvements in traditional performance metrics.

\subsubsection{Developer Study}

\begin{figure*}[t]
\fboxsep=0mm
\includegraphics[width=\linewidth]{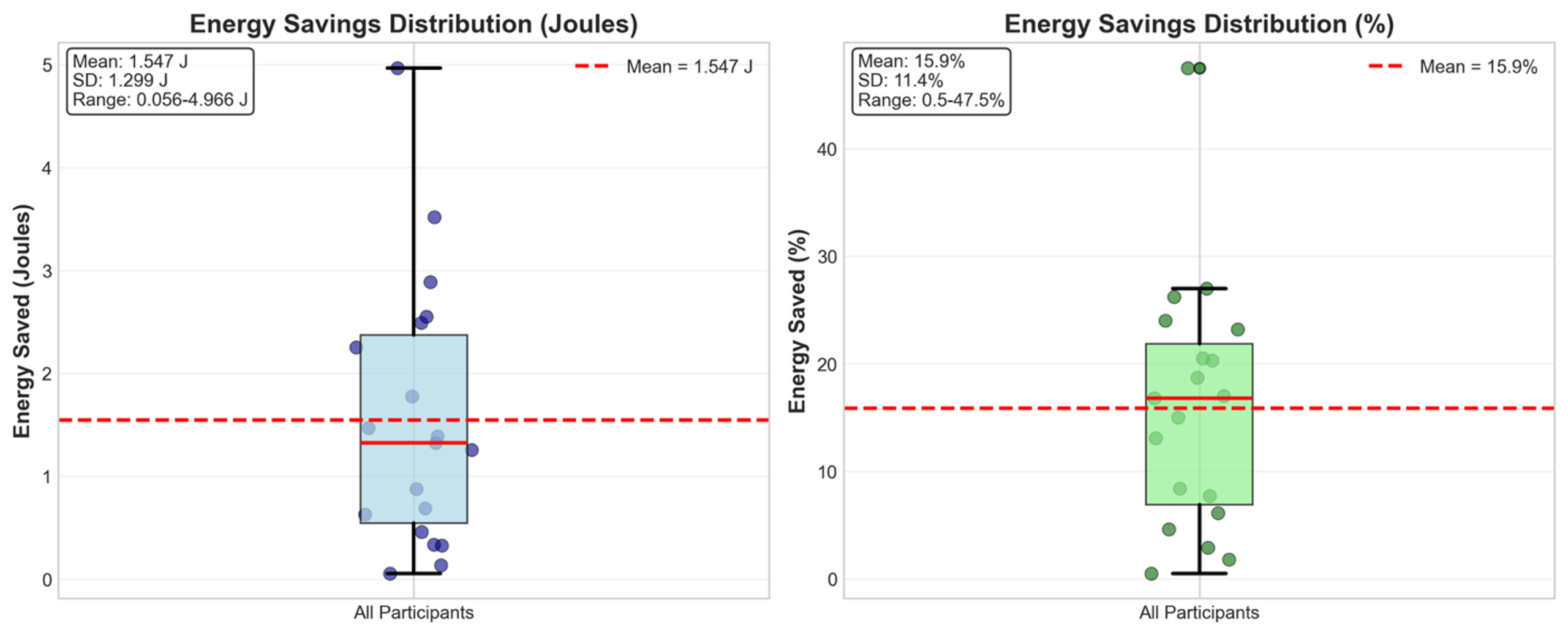}
\caption{Participant-level energy savings with EcoAssist.}
\Description{A pair of box plots displaying participant-level energy savings distribution from the user study, one showing energy savings in joules, the other presenting percentage improvements.}
\label{fig:fig-7}
\end{figure*}

The second component of our evaluation examined how developers engaged with the prototype in practice. Twenty participants used the assistant to generate and optimize a personal portfolio website within the environment, following the workflow described in Section 4.1. Across all twenty cases, the optimizations consistently reduced energy consumption. Savings ranged from 0.06 J to 4.97 J, with an average decrease of 1.55 J corresponding to a 15.9\% improvement in efficiency (Figure \ref{fig:fig-7}). Core functionality was preserved: fifteen participants reported that the optimized page behaved identically to the original, while the remaining five noted that some content was removed or layout changes occurred in input boxes. P10 summarized this experience by noting, ‘it cut down energy use but everything still worked fine,’ underscoring the alignment between efficiency gains and reliable outcomes.

\begin{figure}[t]
\fboxsep=0mm
\includegraphics[width=\linewidth]{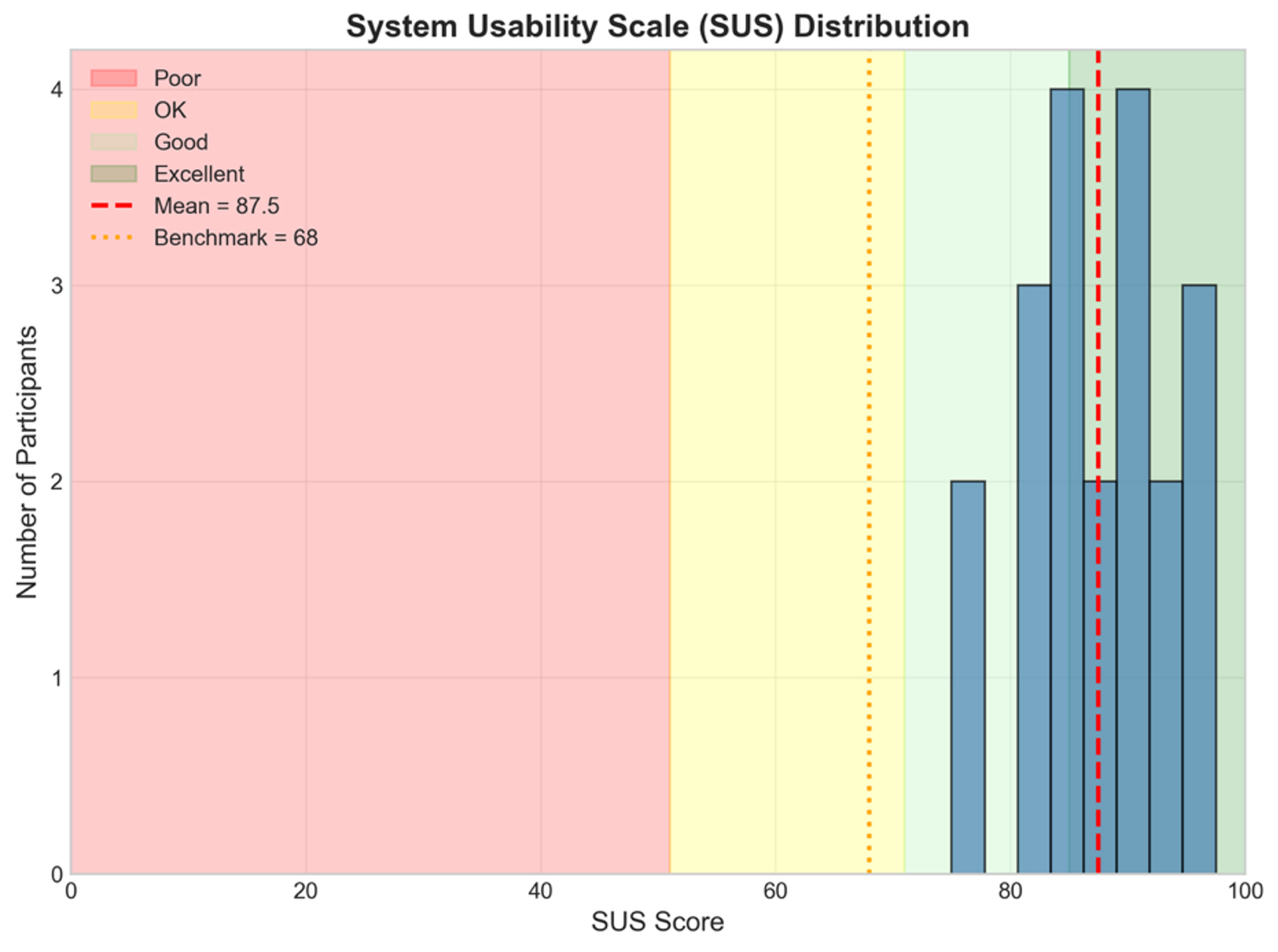}
\caption{SUS score distribution showing excellent usability (M = 89.4).}
\Description{System Usability Scale (SUS) scores from the user study, illustrating that scores lie in the 'excellent usability' range.}
\label{fig:fig-8}
\end{figure}

\begin{figure}[t]
\fboxsep=0mm
\includegraphics[width=\linewidth]{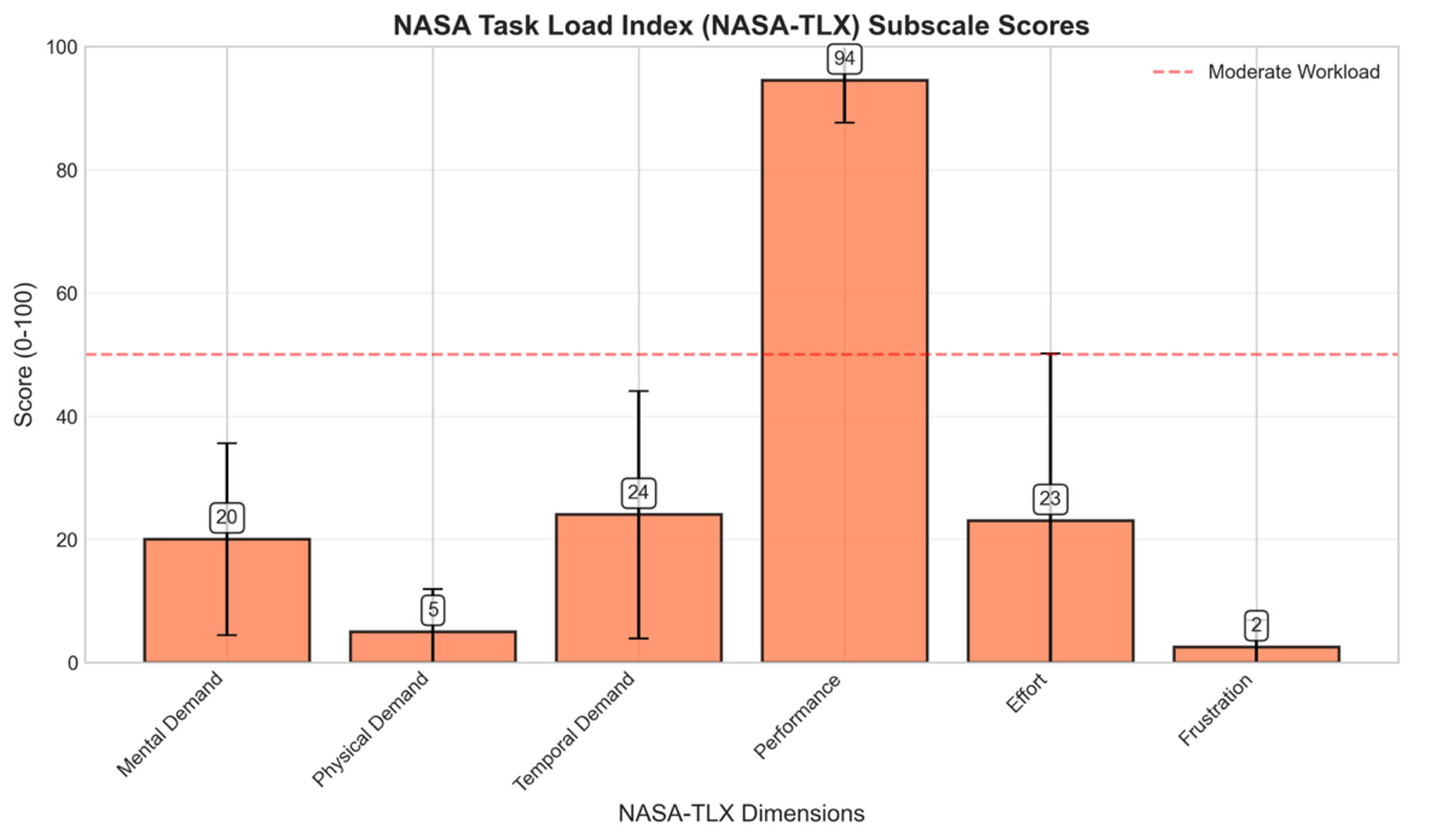}
\caption{NASA-TLX dimension scores indicating low workload across all dimensions.}
\Description{NASA-TLX workload ratings from the user study, including mental demand, physical demand, temporal demand, performance, effort, and frustration.}
\label{fig:fig-9}
\end{figure}

Participants also responded positively to the tool’s usability. The mean System Usability Scale score was $87.5~(SD = 6.02)$, well above the benchmark value of 68 and placing EcoAssist in the ``excellent'' category (Figure \ref{fig:fig-8}). NASA-TLX workload ratings reinforced these impressions, showing very low mental demand ($M = 20.0$), minimal effort ($M = 23.0$), and almost no frustration ($M = 2.5$), with temporal demand also remaining low ($M = 24.0$) (Figure \ref{fig:fig-9}). Participants noted the system fit naturally into their workflow. P2 said, ``it didn’t slow me down at all'', supporting evidence that EcoAssist adds little cognitive burden, key for fast-paced adoption.

\begin{figure}[t]
\fboxsep=0mm
\includegraphics[width=\linewidth]{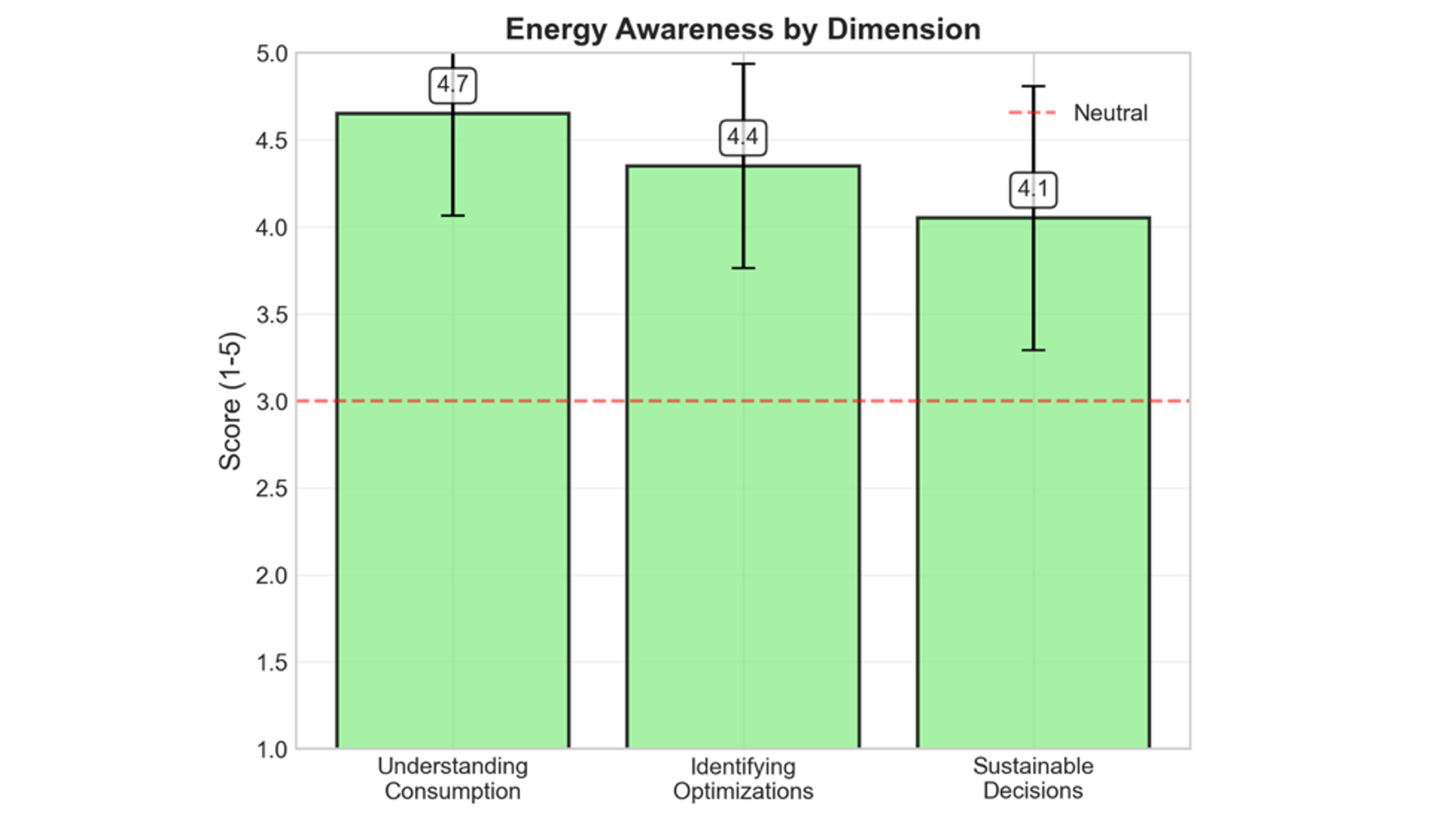}
\caption{Energy awareness scores by dimension.}
\Description{Mean Energy Awareness Scale scores (1–5) across three dimensions: understanding energy consumption, identifying optimization opportunities, and making sustainable decisions.}
\label{fig:fig-10}
\end{figure}

\begin{figure}[t]
\fboxsep=0mm
\includegraphics[width=\linewidth]{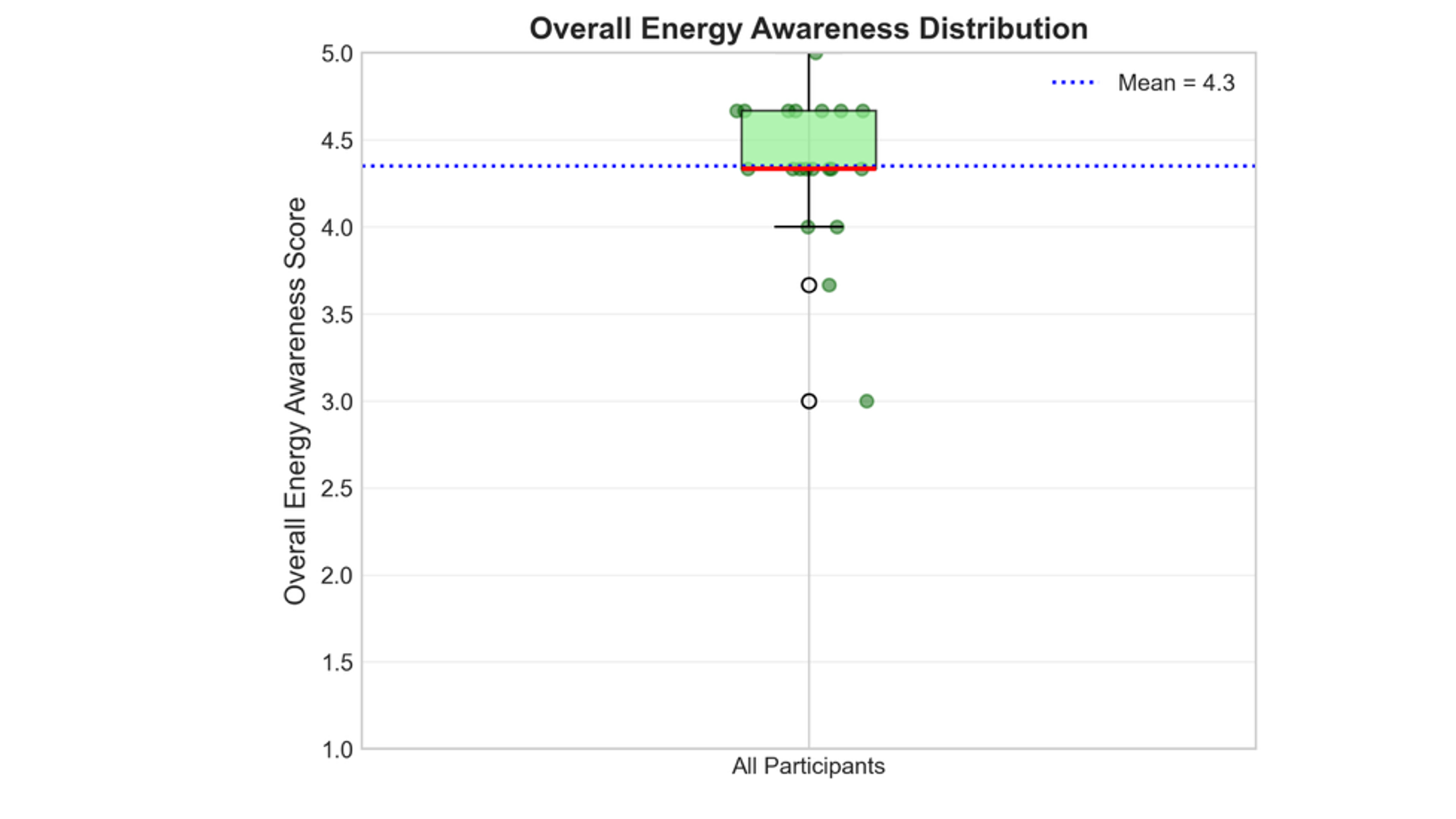}
\caption{Distribution of overall energy awareness scores across participants.}
\Description{Distribution of participants’ overall Energy Awareness Scale scores, shown as a boxplot with individual data points overlaid. The dotted line indicates the mean score. Most responses cluster at the upper end of the scale, with limited dispersion, indicating uniformly high energy awareness among participants.}
\label{fig:fig-11}
\end{figure}

The user study further showed that EcoAssist heightened participants’ awareness of energy impact. The overall Energy Awareness Scale score averaged 4.35 out of 5, with consistently high ratings across dimensions of understanding consumption ($M = 4.65$), identifying optimizations ($M = 4.35$), and making sustainable decisions ($M = 4.05$) (Figures \ref{fig:fig-10} and \ref{fig:fig-11}). Developers often described surprise at the measurable impact of small changes. P4 remarked ``I had no idea this would have any measurable impact,'' while P8 mentioned ``I’ll start paying more attention to page weight and bloat''. At the same time, several participants noted that they don’t usually think about energy during regular development work, with P8 saying they ``never look at page weight unless a site feels heavy.'' Others pointed to team norms as a barrier, with P4 saying they ``would care more about this if our team talked about sustainability at all.'' Additionally, several participants described EcoAssist as a supportive coding partner with an energy focus. As noted by P7, it felt ``like a pair programmer who cares about energy,'' underscoring how the system reframed sustainability as a concrete and immediate concern.

Adoption of EcoAssist’s recommendations was high. On average, participants applied more than 90\% of the suggested edits during the study. Fifteen participants accepted all optimizations without modification, while the remaining five rejected one or two suggestions for personal reasons, such as preferring maximum image quality or small style differences. Participants consistently pointed to the side-by-side diff view as critical for trust: ``I liked seeing exactly what changed, it gave me confidence nothing important was being removed,''. This transparency was important in encouraging acceptance of the system’s optimizations and suggests a viable path for integrating EcoAssist into everyday development workflows.
\section{Discussion}

This paper has explored how AI coding assistants might influence developers’ practices around energy-efficient frontend development. AI tools have unlocked new opportunities for productivity through natural-language programming and rapid iteration, yet our work highlights the persistent challenge of balancing development speed with energy efficiency in developers’ everyday experience. EcoAssist addresses this gap by embedding energy analysis directly into the coding process, helping developers understand and act on energy impact without sacrificing productivity. In this section, we reflect on broader themes that emerged from our evaluation and identify directions for future work.

\subsection{Adoption Challenges}

While our results are promising, deploying a tool like EcoAssist in real-world development environments faces several adoption challenges. A primary hurdle is the prevailing culture and priorities of web development teams: tight deadlines and feature delivery often take precedence over secondary concerns like energy efficiency. Developers in fast-paced workplaces might be reluctant to incorporate an additional step of energy checking unless its benefits are clearly visible and align with their immediate goals. In our study, participants were explicitly motivated to consider energy use, but outside of a research setting many developers may simply ignore or disable an energy-focused tool if they perceive it as slowing them down. This means EcoAssist (or similar assistants) must demonstrate value with minimal friction. For instance, by integrating seamlessly into existing IDEs and continuous integration pipelines, and by providing recommendations that rarely interfere with essential application behavior. Another adoption barrier is awareness and education. Many practitioners are still unfamiliar with sustainable coding principles, so they might not initially recognize why EcoAssist’s suggestions matter. If the tool’s outputs are not intuitively understood (e.g., showing energy saved in joules, which some might find abstract), developers could dismiss the feedback as non-essential. These challenges align with the tendencies expressed by participants in our results.Another challenge lies in incentives. Developers are seldom held accountable for energy efficiency, unlike load time or security, which carry obvious business impact. Without organizational goals, such as eco-efficiency targets, tools like EcoAssist depend on personal initiative. Making the tool easy to try could help adoption, for instance by offering it as a plugin in common editors or as an option alongside mainstream AI assistants like GitHub Copilot. However, establishing its reliability at scale will require testing in real development environments. Ultimately, adoption will hinge not only on technical integration but also on cultural change and awareness within teams. Encouragingly, the industry is placing greater emphasis on sustainability, with new standards and initiatives emerging \cite{greensoftware2024sci,isoiec2024sci,w3c2023wsg}. In this context, EcoAssist could serve as a supportive assistant that helps developers remain productive while producing greener code. Traditional sustainability tools operate after code is written, requiring developers to rerun analyses and re-edit code. By contrast, EcoAssist integrates feedback earlier, embedding checks directly in the AI-assisted coding loop.

\subsection{Integrating Energy Awareness into Developers’ Workflows}

EcoAssist shows that energy feedback can be embedded directly into coding workflows. On one hand, the AI can rapidly identify inefficiencies and suggest optimizations that a developer might miss, especially in a “vibe coding” scenario where the focus is on quick functionality. On the other hand, developers are ultimately responsible for the software and must have control over deciding which suggestions to accept. We designed EcoAssist to support this balance by making its recommendations transparent (through diffs and energy metrics) and leaving control in the user’s hands. The study results indicate that this human-in-the-loop strategy was effective: participants generally trusted the AI’s suggestions, yet they still exercised their own judgment in a few cases where trade-offs arose (e.g., choosing image quality over a small energy gain). This confirms that developers are capable of reasoning about sustainability when given the proper context, and it underscores that the AI’s role is to augment human decision-making, not replace it. The high usability scores (SUS = 89.4) can also be explained by EcoAssist’s effect on productivity. By embedding optimizations directly into the coding process, the tool prevents the productivity loss that typically occurs when developers must stop, run external energy checkers, and then re-edit code. Instead, issues are highlighted as code is generated or reviewed, similar to a spell-checker, which allows developers to address them immediately without breaking focus. This approach turns sustainability checks into part of the natural development flow, reducing context-switching and accelerating iteration. Participants described this integration as supportive rather than burdensome, which helps explain why they were willing to adopt most suggestions. As P5 noted, ‘it felt like the tool was helping me review the code,’ highlighting that EcoAssist’s unobtrusive design was perceived as supporting rather than interrupting developers’ workflow. In practice, we believe this balanced approach is crucial for any AI coding assistant, as it must respect the developer’s goals and knowledge. Nevertheless, our study offers only an initial view of developers’ attitudes, and further work is needed to understand why sustainability-oriented tools are adopted, deprioritized, or abandoned in everyday projects.

Benchmark results show that EcoAssist can reduce energy use by intervening where web code consumes the most resources: scripts, images, and load-time execution. Because the model was trained on before-and-after code pairs with significant energy improvements, it learned not only which transformations save energy but also when they are most effective. For instance, minifying and eliminating unused CSS and JavaScript reduces the cycles needed for parsing and execution, lowering processor energy use. Furthermore, transcoding images into modern formats and compressing them at appropriate resolutions cuts network transfer size and decoding cost, reducing energy on both servers and client devices. Additionally, extracting and inlining critical-path CSS and subsetting web fonts remove render-blocking requests and shrink font payloads, cutting early CPU wake-ups, style recalculation, and network-transfer energy during initial rendering. Each of these interventions eliminates avoidable power expenditure at different stages of page processing, and together they account for the consistent 15–20 percent energy reductions observed in our benchmark, particularly for applications with substantial size and JavaScript content that mirror real-world websites.

We found that presenting concrete metrics, such as the joules saved, helped developers appreciate why the AI’s suggestion was useful. Some participants suggested that richer dialogue could make this balance even stronger. For example, a developer might ask why a particular optimization was recommended, and the assistant could explain that compressing an image makes the page load faster and use less energy. Exchanges like this would help developers stay more informed and in control. In this setup, the assistant focuses on sustainability while the developer considers the bigger picture. Striking the right balance means the AI should be assertive enough to flag significant issues (so that inefficiencies aren’t overlooked) but leaves the final choice to the developer. Our study shows this balance is possible, as participants did not feel overruled but saw the tool as offering helpful suggestions. This integration helped participants see sustainability as relevant without feeling slowed down.

\subsection{Limitations \& Future Work}

While EcoAssist’s results are encouraging, several limitations in our work open avenues for future research. First, our evaluation focused on relatively self-contained frontend pages. Real-world web applications, for example, large e-commerce sites or complex single-page applications built with frameworks like React/Angular, may present new challenges. Although EcoAssist was not trained on these patterns, its techniques, such as minification, image compression, and lazy loading, are expected to carry over effectively. Integrating and deploying EcoAssist within modern build tools and frameworks is a logical next step to ensure it can handle the full complexity of contemporary web development. Another limitation is that our energy measurements were conducted in a controlled setting (using a specific hardware/software setup with Powermetrics \cite{apple2023powermetrics} on macOS). This approach enables us to compare results consistently. However, the exact amount of energy saved in real-world use will vary depending on the device, browser, and network conditions. Our measurements were conducted on an ARM-based MacBook Air M4, and savings may differ on x86 servers or other client devices, motivating cross-platform validation in future work. Future studies could therefore test EcoAssist across a broader range of hardware and use standardized metrics, such as the Software Carbon Intensity \cite{greensoftware2024sci} (SCI), to express savings in carbon-equivalent terms. This would make it easier to communicate and compare the environmental impact in a more universal way.

There is also room to improve the breadth and intelligence of optimizations. Our fine-tuned model was trained on specific patterns of inefficiency that we identified and measured in the training corpus. It performs well on those, but it might miss opportunities outside its training distribution. In the future, incorporating a broader set of data, including more diverse websites or even mobile app frontends, could teach the model to recognize additional inefficiency patterns. Another intriguing direction is to explore proactive energy-aware code generation. In our current workflow, the AI assistant (e.g., GPT-5) generates code without considering energy, and then EcoAssist optimizes it. An alternative (or complementary) approach would be to integrate energy awareness directly into the code generation phase. If the next generation of coding assistants could be nudged to avoid certain anti-patterns from the start, the role of a tool like EcoAssist might evolve more towards validation and fine-tuning rather than heavy correction. 

From a human-centered perspective, a limitation of our user study is its scope and duration. While participants’ feedback was positive, real adoption would involve long-term use and exposure to a variety of projects. Future studies could deploy EcoAssist in a team’s day-to-day workflow over several weeks or months to observe sustained usage patterns, potential fatigue, or more profound learning effects. It would be valuable to see if developers continue to use the tool when not observed, how it influences the quality of code reviews or team conventions around sustainability, and whether it remains helpful in day-to-day work.

Another limitation is our measure of energy awareness. We designed a short, study-specific scale inspired by prior work on developer awareness \cite{jagroep2017awakening}. While this captured initial reactions, it does not fully represent developers’ attitudes, motivations, or the organizational factors that shape adoption of energy-aware tools. Our quantitative evaluation used SUS and NASA-TLX, which assess usability and workload but offer limited insight into adoption barriers such as perceived relevance or workflow pressures. Future work should build a richer, validated measurement strategy that pairs energy-awareness metrics with attitudinal and longitudinal measures, including intention to adopt, perceived workflow fit, and continued use over time, to better understand how tools like EcoAssist influence practice.

A further consideration is the environmental footprint of the GPT-4o-mini model that powers EcoAssist. Independent benchmarks report $3.098 \pm 0.639~\mathrm{Wh}$ per long-prompt inference ($\approx 10$k input tokens) for GPT-4o-mini on A100-class hardware~\cite{jegham2025hungry}. OpenAI has not released fine-tuning energy for GPT-4o-mini, so we reference measured values from architecturally similar 7--9B decoder models. For instance, fine-tuning a LLaMA-family 7--9B decoder model with a parameter-efficient method has been measured to consume approximately $0.485~\mathrm{kWh}$, based on recent evaluations of LLaMA-3.1-8B~\cite{pingua2025medical}. During typical use, developers make multiple EcoAssist requests while iterating on a website. Assuming roughly 200 long-prompt requests, the inference cost is about $0.6~\mathrm{kWh}$ ($200 \times 3.098~\mathrm{Wh}$). Combined with the one-time fine-tuning proxy ($0.485~\mathrm{kWh}$), this yields an estimated overhead of $\approx 1.1~\mathrm{kWh}$. With EcoAssist reducing frontend energy use by 13--16\%, this overhead is recovered once the underlying frontend would have consumed roughly 7--9~$\mathrm{kWh}$ (using a standard breakeven estimate based on the reduction rate). Excluding the one-time fine-tuning proxy, the breakeven falls to roughly 4--5~$\mathrm{kWh}$. Although approximate, this estimate suggests that EcoAssist can provide energy savings under modest real-world workloads. For context, the Web Energy Archive reports that websites typically consume 10--200~$\mathrm{Wh}$ per 1{,}000 page views~\cite{philippot2014characterization}. Using a mid-range intensity of 50~$\mathrm{Wh}$ per 1{,}000 views, the 4--5~$\mathrm{kWh}$ breakeven corresponds to approximately 80k--100k page views, and the 7--9~$\mathrm{kWh}$ breakeven to roughly 140k--180k views. Empirical traffic reports show monthly page-view volumes ranging from a few thousand to several hundred thousand per site~\cite{angelou2024social,xavier2024web}. On this basis, assuming a conservative scale of $\sim$50k page views per month for a moderately active production website, these breakeven thresholds (80k--180k views) correspond to roughly 2--4 months of typical traffic. Refining these values will depend on future access to model-level energy data. Future iterations could examine whether smaller model variants can provide similar guidance while reducing inference cost, making the assistant more practical for everyday use and easier to deploy in real development settings.

Overall, EcoAssist shows how energy awareness can become part of the everyday experience of AI-assisted coding. Addressing these limitations and exploring the suggested future directions will be crucial to fully realizing the vision of sustainable-by-default coding workflows.
\section{Conclusion}

Our work bridges emerging efforts in sustainable web development with the rapid rise of AI coding assistants, offering a novel approach to the overlooked problem of energy efficiency in frontend code. Through a formative study and system design, we identify key shortcomings in how current AI-assisted workflows neglect energy efficiency concerns. In response, we develop EcoAssist, an energy aware coding assistant that analyzes AI-generated frontend code, highlights energy-intensive patterns, and suggests optimized alternatives. Our evaluation demonstrates that developers using EcoAssist are more likely to produce efficient, lower-energy code without sacrificing functionality or speed. By embedding energy feedback cues directly into everyday development workflows, this work represents a first step toward fostering energy-conscious practices in human–AI collaborative frontend development.

\begin{acks}
This work was supported by ITI/LARSyS through Funda\c{c}\~ao para a Ci\^encia e a Tecnologia (FCT) funding (projects LA/P/0083/2020 and UID/50009/2025) and by the LogaCulture project (Grant Agreement No. 101094036). The work of Nikolas Martelaro was partially supported by Portugal's Funda\c{c}\~ao para a Ci\^encia e a Tecnologia (FCT) through the Carnegie Mellon University Portugal Program.
\end{acks}
\balance
\bibliographystyle{ACM-Reference-Format}
\bibliography{References/CHI2025_Work_Vision}

\appendix 

\clearpage
\onecolumn
\section{APPENDICES}

\subsection{EcoAssist Evaluation User Study Guide}

\subsubsection*{EcoAssist Evaluation}

\textbf{Duration:} $\sim$45 minutes per participant
\textbf{Format:} In-person, laptop-based coding tasks with EcoAssist prototype

\subsubsection*{HCI Research Question}

How can sustainability-aware coding assistants help developers integrate energy efficiency into frontend workflows, making trade-offs visible while preserving functionality and usability, without reducing development speed?

\subsubsection*{Setup}

\begin{itemize}
    \item Laptop running EcoAssist prototype
    \item Screen and audio recording enabled
    \item Quiet environment
\end{itemize}

\subsubsection*{Participants}

\begin{itemize}
    \item 20 developers
    \item All fill out a short intake form (background, coding experience, sustainability familiarity)
\end{itemize}

\subsubsection*{Facilitation Guidelines}

\begin{itemize}
    \item Reassure participants that the tool is being evaluated, not them
    \item Encourage think-aloud behavior
\end{itemize}

\subsubsection*{Participant Instructions}

``You’ll be using a coding assistant called EcoAssist. The goal is to observe how developers interact with it during typical frontend coding tasks. This is not a test of your skills, we are evaluating the tool itself. There are no right or wrong answers, and you can use however it feels natural to you. Please try to think out loud as you go, say what you’re doing or noticing, and feel free to comment on anything confusing, interesting, or unexpected.''

\subsubsection*{Evaluation Flow}

\textbf{Total Session Time:} $\sim$45 minutes

\begin{enumerate}
    \item Welcome \& Consent -- 5 minutes
    \item Pre-Study Form -- 5 minutes
    \item Coding Task -- 15--20 minutes
    \item Post-Study Questionnaire \& Interview -- 10 minutes
\end{enumerate}

\subsubsection*{Consent Form}
Name: \rule{4cm}{0.4pt} \hspace{1cm} Signature: \rule{4cm}{0.4pt}

I understand that:
\begin{itemize}
    \item This session will be recorded (screen + audio)
    \item My data will be anonymized and used for research only
    \item I can stop at any time without penalty
\end{itemize}

Date: \rule{3cm}{0.4pt}

\subsection*{Pre-Study Questions}

\textbf{1. Role}

$\square$ Student Developer \hspace{0.5cm}
$\square$ Professional Developer \hspace{0.5cm}
$\square$ Other

\noindent\textbf{2. Experience}

\begin{itemize}
    \item Years of coding experience: \rule{3cm}{0.4pt}
    \item Years using frontend frameworks (e.g., React, Vue, Angular): \rule{3cm}{0.4pt}
    \item Used AI coding assistants before (e.g., Copilot, Cursor, Claude Code)?
    
    $\square$ Yes \hspace{0.5cm}
    $\square$ No
\end{itemize}

\noindent\textbf{3. Sustainability Background}

\begin{itemize}
    \item Familiar with sustainable web development practices?

    $\square$ Not at all \hspace{0.5cm}
    $\square$ A little \hspace{0.5cm}
    $\square$ Somewhat \hspace{0.5cm}
    $\square$ Very familiar

    \item Have you used tools that measure energy or emissions in code?

    $\square$ Yes \hspace{0.5cm}
    $\square$ No \hspace{0.5cm}
    $\square$ Not sure
\end{itemize}

\subsection*{Coding Task -- EcoAssist}

\textbf{Actions:}

\begin{enumerate}
    \item Use EcoAssist AI assistant (GPT-5) to generate a portfolio app.
    \item Observe the sustainability feedback provided by the tool.
    \item Optimize the generated code.
    \item Review the optimized metrics.
    \item Verify that the optimized app still functions correctly using the preview button.
\end{enumerate}

\noindent\textbf{Questions for each website update:}\\
\textbf{1.} How much energy did EcoAssist save compared to the original code? (in Joules)

\textit{Short answer text}

\noindent\textbf{2.} Did the optimized app preserve the original behavior/functionality?

\begin{itemize}
    \item Yes, everything worked as expected
    \item Mostly, with small differences
    \item No, some functionality was broken
\end{itemize}

\noindent\textbf{3.} Did you reject some of the suggested optimizations? If yes, why?

\subsection*{Post-Study Questionnaire}

\textbf{Usability (Linear Scale (1--5))}

\begin{enumerate}
    \item I think that I would like to use this system frequently.
    \item I found the system unnecessarily complex.
    \item I thought the system was easy to use.
    \item I think that I would need the support of a technical person to be able to use this system.
    \item I found the various functions in this system were well integrated.
    \item I thought there was too much inconsistency in this system.
    \item I would imagine that most people would learn to use this system very quickly.
    \item I found the system very cumbersome to use.
    \item I felt very confident using the system.
    \item I needed to learn a lot of things before I could get going with this system.
\end{enumerate}

\noindent\textbf{Workload (0--100 (Low $\rightarrow$ High))}

\begin{enumerate}
    \item Mental Demand: ``How mentally demanding was the task?''
    \item Physical Demand: ``How physically demanding was the task?''
    \item Temporal Demand: ``How hurried or rushed was the pace of the task?''
    \item Performance: ``How successful were you in accomplishing what you were asked to do?''
    \item Effort: ``How hard did you have to work to accomplish your level of performance?''
    \item Frustration: ``How insecure, discouraged, irritated, stressed, and annoyed were you?''
\end{enumerate}

\noindent\textbf{Energy Awareness (Linear Scale (1--5))}

\begin{enumerate}
    \item The software helped me understand my code’s energy consumption.
    \item The software supported me in identifying ways to optimize energy use.
    \item The software increased my awareness of sustainable energy practices and decision-making for software development.
\end{enumerate}

\end{document}